# Nanoscale

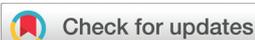 Check for updates

## Citrate stabilized gold nanoparticles interfere with amyloid fibril formation: D76N and ΔN6 β2-microglobulin variants†

Giorgia Brancolini, [ID] *[a] Maria Celeste Maschio, [ID] [a] Cristina Cantarutti,[b] Alessandra Corazza,[b,c] Federico Fogolari, [ID] [b,c] Vittorio Bellotti,[d,c,e] Stefano Corni[f] and Gennaro Esposito [ID] *[a,c,g]

Protein aggregation including the formation of dimers and multimers in solution, underlies an array of human diseases such as systemic amyloidosis which is a fatal disease caused by misfolding of native globular proteins damaging the structure and function of affected organs. Different kind of interactors can interfere with the formation of protein dimers and multimers in solution. A very special class of interactors are nanoparticles thanks to the extremely efficient extension of their interaction surface. In particular citrate-coated gold nanoparticles (cit-AuNPs) were recently investigated with amyloidogenic protein β2-microglobulin (β₂m). Here we present the computational studies on two challenging models known for their enhanced amyloidogenic propensity, namely ΔN6 and D76N β₂m naturally occurring variants, and disclose the role of cit-AuNPs on their fibrillogenesis. The proposed interaction mechanism lies in the interference of the cit-AuNPs with the protein dimers at the early stages of aggregation, that induces dimer disassembling. As a consequence, natural fibril formation can be inhibited. Relying on the comparison between atomistic simulations at multiple levels (enhanced sampling molecular dynamics and Brownian dynamics) and protein structural characterisation by NMR, we demonstrate that the cit-AuNPs interactors are able to inhibit protein dimer assembling. As a consequence, the natural fibril formation is also inhibited, as found in experiment.

## Introduction

The interest in the interaction of nanoparticles (NPs) with amyloidogenic proteins is continuously growing due to the huge number of possible applications in nanomedicine and nanotechnology.[1–3] In particular, the interaction between gold nanoparticles (AuNPs) and the biological systems has received great attention due to the development of novel therapeutic and diagnostic tools,[4,5] and due to concerns regarding their safety *in vivo*.[6,7] It is widely accepted that the contact between the surface of NPs and proteins triggers a competition between different biological molecules to adsorb on the surface of the NPs[5] either transiently or permanently, in the so-called soft or hard corona layer.[8] As a consequence, the protein structure and/or function may be perturbed to different extent or remain conserved.

[a]Center S3, CNR Institute Nanoscience, Via Campi 213/A, 41125 Modena, Italy.
E-mail: giorgia.brancolini@nano.cnr.it
[b]Dipartimento di Scienza Mediche e Biologiche (DSMB), University of Udine,
Piazzale Kolbe 3, 33100 Udine, Italy
[c]Istituto Nazionale Biostrutture e Biosistemi, Viale medaglie d'Oro,
305 - 00136 Roma, Italy
[d]Dipartimento di Medicina Molecolare, Universita' di Pavia, Via Taramelli 3,
27100 Pavia, Italy
[e]Division of Medicine, University College of London, London NW3 2PF, UK
[f]Department of Chemical Science, University of Padova, via VIII Febbraio 2,
35122 Padova and Center S3, CNR Institute Nanoscience, Via Campi 213/A,
41125 Modena, Italy
[g]Science and Math Division, New York University at Abu Dhabi, Abu Dhabi,
United Arab Emirates. E-mail: gennaro.esposito@uniud.it
†Electronic supplementary information (ESI) available. See DOI: 10.1039/
C7NR06808E

Understanding protein–inorganic nanoparticle interactions is central to the rational design of new tools in biomaterial sciences, nanobiotechnology and nanomedicine. Theoretical modelling and simulations provide complementary approaches for experimental studies.

We have recently studied the interaction of citrate-capped gold nanoparticles (cit-AuNPs) with β2-microglobulin (β₂m),[9] the light chain component of class I major histocompatibility complex (MHCI), see Fig. 1. In long-term hemodialysed patients, this protein precipitates into amyloid deposits and accumulates in the collagen-rich tissues of the joints, originating a pathology referred to as dialysis related amyloidosis (DRA).[10] Contrary to expectations, based on previous studies of

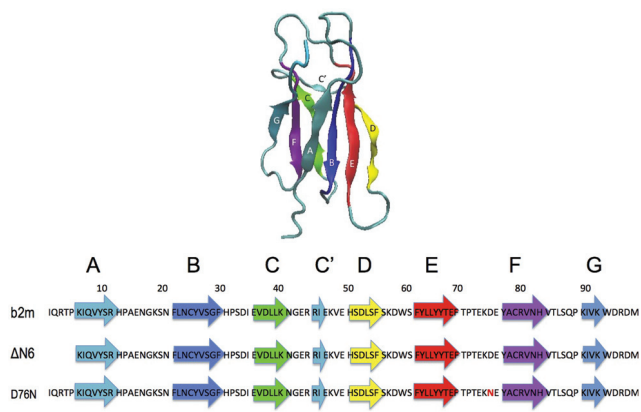

**Fig. 1** The native structure of wild-type human β-2 microglobulin (β2m) (top) and its secondary structure content together with that of ΔN6 and D76N β2m variants (bottom).

β2m with various nanoparticle systems of larger sizes,[11] no clear aggregation promotion and/or inhibition was detected, in the presence of citrate-coated AuNPs with diameter of 5 nm.[9]

Here, we progress further the investigation of nanoparticle effects on more challenging amyloidogenic β2m protein species, namely D76N and ΔN6 that can undergo fibrillogenesis under mild conditions at neutral pH. D76N is a naturally occurring variant of β2m bearing an asparagine residue at position 76 instead of an aspartate. This single point mutant ASP76ASN (D76N) is associated with the late onset of a fatal hereditary systemic amyloidosis characterised by extensive visceral amyloid deposits. This variant readily forms fibrils by agitation at neutral pH exhibiting the highest amyloidogenic ability amongst all known β2m variants.[12]

The ΔN6 is a truncated form of β2m, lacking the first six N-terminal residues. This cleaved variant is the major component of *ex vivo* amyloid plaques (~26%) of patients affected by DRA.[13] While there is a broad agreement regarding the ability of ΔN6 to prime the fibrillar conversion of Wild-Type β2m *in vitro* under physiological conditions, the mechanism by which this occurs is not consensual. Notwithstanding that a prion-like mechanism of ΔN6 has been proposed to drive the fibrillogenesis of the β2m native form,[14] Bellotti and coworkers has challenged the prion-like hypothesis by reporting that the Wild-Type β2m does not fibrillate with monomeric ΔN6 but rather with preassembled fibrils of ΔN6.[15]

The major goal of the present work is to address the interaction mechanism between cit-AuNPs and D76N and ΔN6 adducts *via* enhanced molecular dynamics and NMR experiments. The focus is placed on the interference of the cit-AuNPs with the protein at the early stages of aggregation, namely monomeric and dimeric adducts.

By using molecular simulations at multiple levels (enhanced sampling molecular dynamics and Brownian dynamics) we provide a map of the preferential interaction sites between monomeric and dimeric protein aggregates and the cit-AuNP. The achieved results on D76N demonstrate that simulations and NMR data provide a picture in which the interaction with cit-AuNP occurs *via* protein dimers, suggesting the presence of preassembled D76N dimers in solution at neutral pH. For ΔN6 variant, preferential interactions are mostly occurring through the amino-terminal region in both the monomeric and dimeric species. At physiological pH, ΔN6 variant may be present as monomers and/or dimers in solution and the interaction with the cit-AuNPs may occur with both, indistinctly.

In all cases, this binding of nanoparticles is able to block the active sites of protein domains used for the binding with another protein, thus leading to an inhibition of the fibrillation activity as found in experiments.

## Results and discussion

The nature of the binding of D76N and ΔN6 variants on cit-AuNP, is characterized by a comprehensive multiple level modeling investigation, spanning from rigid-body protein–surface docking to enhanced molecular dynamics simulations. In this section we describe the employed computational approach and the results obtained for the monomeric and dimeric adducts. In the Experimental part we will report the experimental NMR and UV-vis data and the comparison with simulations.

Brownian dynamics (BD) simulations are initially performed to generate protein–surface monomeric and/or dimeric encounter complexes, by keeping the internal structure of the proteins and the surface rigid during the docking. More specifically, the adsorption free energies of the encounter complexes are computed for the structures resulting from the docking and the BD simulation trajectory are clustered to identify different orientations. For each of the most populated complexes, which are ranked by size, a representative structure is selected for each system and refined by enhanced MD.

The BD interaction energy of the protein with the cit-AuNP surface is described by four main terms:[20] van der Waals energy described by site–site Lennard-Jones interactions, $E_{LJ}$, adsorbate–metal electrostatic interaction energy, $U_{EP}$ and the desolvation energy of the protein, $U_{ds}^p$, and of the metal surface, $U_{ds}^m$.[21]

For the monomeric assemblies, the simulations are started from the NMR structure (PDB:1JNJ) upon inclusion of variations, whereas for the dimeric assemblies a preliminary protein–protein dockings is also performed to obtain the initial more favourable association complexes (results are reported in section "Protein–protein docking: the dimeric interface").

After performing the initial docking simulations, the stability of the docked encounter complexes is assessed by running Replica-Exchange simulations in solvent and on cit-AuNP involving multiple simulations at different temperatures (T-REMD). The adopted simulation protocol[9] includes 20 (or 30) ns of replica exchange molecular dynamics (REMD) at

## D76N monomer

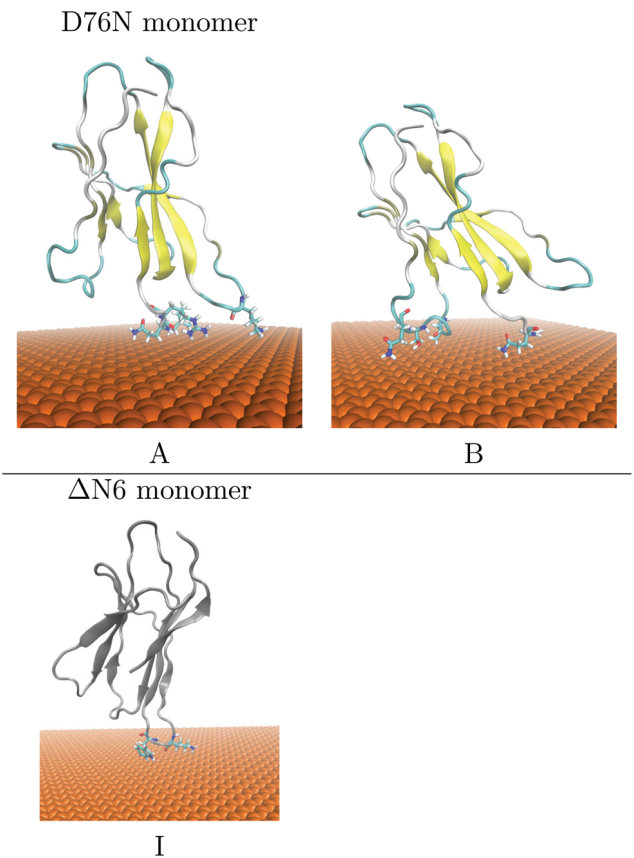

A

B

## ΔN6 monomer

I

**Fig. 2** Most populated encounter complexes of $D76N_{chg}^{net} = +3.00e$ and $\Delta N6_{chg}^{net} = -1.00e$ on negatively charged gold ($Au_{chg}^{neg} = -0.05e$) obtained by BD simulation. Complexes A, B are the representative structures of the two clusters obtained for D76N (including 71 and 29% of the orientations, respectively). Complex I is the representative complex of the most populated cluster identified for ΔN6 (including the 96% of the protein–surface reciprocal orientations). The protein backbone is shown in cartoon representation (with the yellow colour for D76N and gray color for ΔN6). The residues contacting the gold surface are shown in stick representation.

different temperatures (T-REMD), yielding an aggregated simulation time of 640 (or 960) ns.

### Monomers on cit-AuNP

**Docking of monomers on negative gold.** In this section, we focus on the docking of D76N and ΔN6 monomers on cit-AuNP. The density of negative charge of the gold surface atoms is chosen according to an atomistic model which is able to mimic the electrochemical potential of the cit-AuNPs surface under aqueous conditions and at physiological pH.[9,22–24]

After this docking procedure was applied to $D76N_{chg}^{net} = +3.00e$ ($X_{chg}^{net}$ = total net charge of $X$ species) and $\Delta N6_{chg}^{net} = -1.00e$ monomers on negative charged gold surface atoms ($Au_{chg}^{net} = -0.05e$), a hierarchical clustering algorithm (based on a minimum distance linkage function) was applied to the diffusional encounter complexes. Two main orientations are found namely A and B, accounting for 71% and 29% of the total encounter complexes, respectively. In the case of ΔN6 monomer, docking provided a single orientation i.e. complex I, accounting for the 96% of the total encounter complexes. Protonation state of the proteins is determined as explained in the Methodology.

The representative structures of the resulting complexes are shown in Fig. 2. The complexes stability and the protein residues contacting the surface are listed in Table 1.

From Table 1, the binding of D76N on cit-AuNP in complexes A and B is driven mostly by the electrostatic terms. The binding in complex A and B is stabilised mostly by the electrostatic terms. The preferred orientation involves the residues at the N-terminal (ILE1 GLN2 ARG3) tail and DE-loop (LYS58). The strong and highly populated binding seems to be associated with the total charge of the gold surface atoms and the amount of charged residues (ARG3, LYS58) contacting the surface and this is due to the fact that in presence of negatively charged gold the protein is able to use simultaneously more than one charged contact in order to optimise the binding. On the contrary, binding of ΔN6 on cit-AuNP in complex I is driven by $E_{LJ}$ interactions but electrostatic is also relevant due to the charge contact of the negative surface with a positively charged LYS58.

**Table 1** Encounter complex from rigid-body BD docking of $D76N_{chg}^{net} = +3.00e$ and $\Delta N6_{chg}^{net} = -1.00e$ (obtained from PDB:1JNJ and modificated, truncated manually) to an Au (111) surface

| D76N Label | Monomer RelPop %[a] | $U_{Repr}$[b] | $E_{LJ}$[c] | $E_{LJ} + U_{ds}^{p} + U_{ds}^{m}$[d] | $U_{EP}$[e] | Contact residues[f] |
|---|---|---|---|---|---|---|
| A | 71 | −65.873 | −2.565 | 17.201 | −83.074 | ILE1 GLN2 ARG3 LYS58 |
| B | 29 | −63.623 | 4.942 | 15.376 | −79.001 | GLN2 VAL85 SER88 GLN89 |

| ΔN6 Label | Monomer RelPop %[a] | $U_{Repr}$[b] | $E_{LJ}$[c] | $E_{LJ} + U_{ds}^{p} + U_{ds}^{m}$[d] | $U_{EP}$[e] | Contact residues[f] |
|---|---|---|---|---|---|---|
| I | 96 | −10.030 | −18.200 | 4.436 | −17.220 | LYS58 TRP60 |

[a] Relative population of this cluster. [b] $U_{Repr}$: total interaction energy of the representative of the given cluster in kT with $T = 300$ K. [c] $E_{LJ}$: Lennard-Jones energy term for the representative complex, in kT. [d] $U_{ds}^{p}$: non-polar (hydrophobic) desolvation energy of the representative complex, in kT. [d] $U_{ds}^{m}$: surface desolvation energy of the representative complex, in kT. [e] $U_{EP}$: total electrostatic energy of the representative complex, in kT. [f] Residues with atoms contacting gold at distances ≤3 Å.

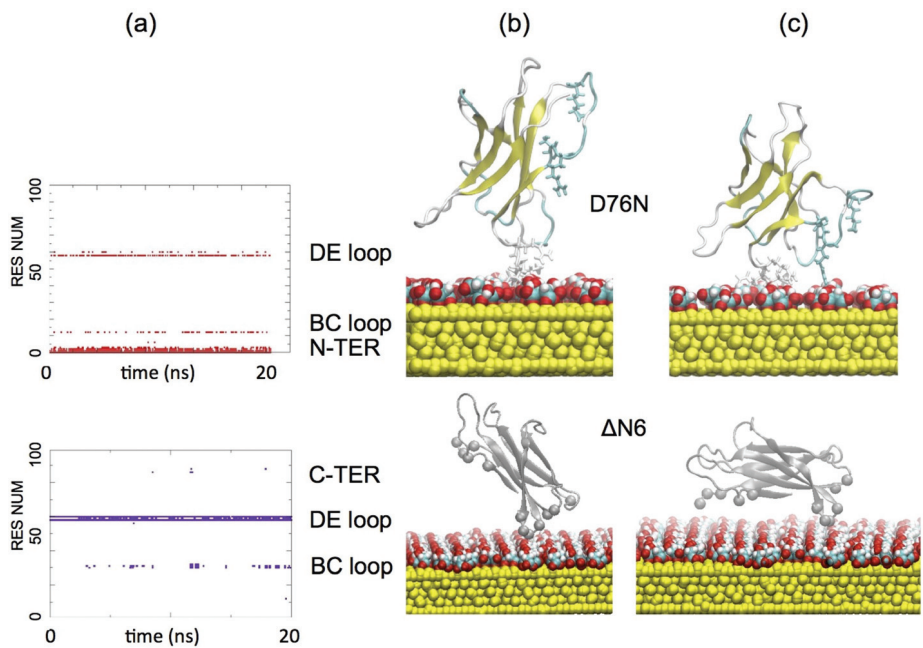

**Fig. 3** Top panel. On the left: Time evolution of contacting residues for the monomeric D76N with respect to the surface of the nanoparticle interface (*i.e.* protein residues within 3 Å from the surface), extracted from the total 20 ns T-REMD and central and right panels report the two most representative structures of the D76N monomer during T-REMD. Bottom panel. ΔN6 binding to citrate-auNP is conserved during the entire 20 ns length of T-REMD since the protein remains anchored through the DE-loop residues (LYS58, TRP60) and BC-loop residue (HIS31). In addition, the ΔN6 monomeric protein exhibited few contacts with C-TER (ARG97, MET99) residues in the very last part of the 20 ns simulation. The capability of the ΔN6 protein to remain anchored to the citrate surface during T-REMD is in line with the intensity reduction which were observed experimentally for ΔN6 on citAuNP.

From the present docking results, we may conclude that the effect of two variations located in significantly different protein domains *i.e.* N-TER for ΔN6 and EF-loop for D76N, does not significantly affect the global orientation of protein bound complexes to cit-AuNP respect to the native β₂m protein. The most populated A and I complexes of the two variants are contacting cit-AuNP through DE-loop (LYS58).

**Enhanced sampling of monomers on cit-AuNP.** To assess the stability of the monomeric docked encounter complexes and to include the effect of structural relaxation, Replica-Exchange simulations in solvent and on cit-AuNP involving multiple simulations at different temperatures (T-REMD) are performed starting from the most representatives and populated monomeric complexes obtained from rigid-body BD docking.

Simulation results of Complex A for D76N monomer and Complex I for ΔN6 interacting with cit-AuNP are summarised in Fig. 3. Panels (a) report time evolution of D76N contacting residues on cit-AuNP, panels (b) and (c) report the final representative structures of the two most recurrent orientations found on the D76N and ΔN6 variants interacting with cit-AuNP. For the ΔN6 variant a longer simulation time of 30 ns was required in order to perform a satisfactory sampling of the contacting patches with the negatively charged citrate layer. Orientations are obtained following the replica at the lowest temperature during the 20 ns and 30 ns of T-REMD, respectively.

**Table 2** Most populated encounter complex for D76N and ΔN6 protein–protein complexes by BD simulation. The structure of a single complex is representative for the 97% of the total encounter complexes for D76N dimers, whereas the ΔN6 dimers is representative for the 70% of the total complexes. The protein backbone is shown in cartoon representation. For nomenclature, see Fig. 1

| Label | RelPop %[a] | $U_{Repr}$[b] | $U_{ds}^p$[c] | $U_{EP}$[d] | Spread[e] | Representative of cluster[f] |
|---|---|---|---|---|---|---|
| D76N | | | | | | |
| A | 97 | −21.821 | −15.287 | −11.624 | 1.462 | |
| ΔN6 | | | | | | |
| I | 70 | −8.693 | −9.071 | −0.162 | 6.6 | |
| J | 30 | −7.585 | −8.627 | 1.551 | 2.7 | |

[a] Relative population of this cluster. [b] $U_{Repr}$: total interaction energy of the representative of the given cluster in kT with T = 300 K. [c] $U_{ds}^p$: non-polar (hydrophobic) desolvation energy of the representative complex, in kT. [d] $U_{EP}$: total electrostatic energy of the representative complex, in kT. [e] RMSD of the structures within the cluster with respect to the representative complex. [f] Representative of a given cluster.

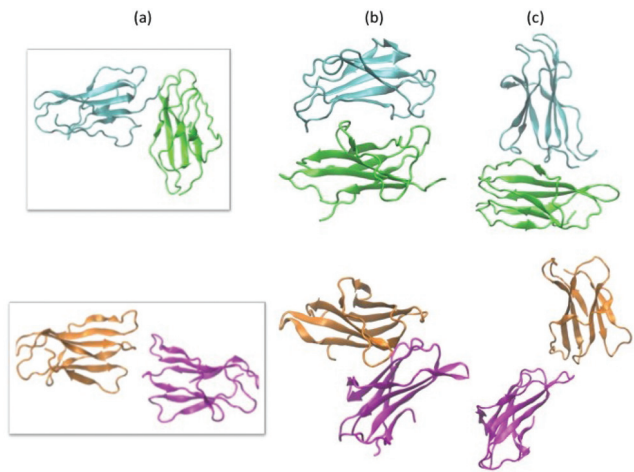

**Fig. 4** Reciprocal orientations of two identical D76N and ΔN6 proteins within dimers obtained starting from the docking after 400 ns MD refinement at 300 K. A series of four simulations were performed on the initial complex obtained from docking, each with different initial velocities. (a) Initial orientation from rigid docking, (b) and (c) most stable zipped and unzipped final orientation after MD refinement of D76N (green and cyan) and ΔN6 (orange and mangenta), respectively.

For D76N mutant, (see upper panels of Fig. 3), a stable interaction between the N-terminal, BC and DE loop and the NP surface is confirmed by T-REMD, but also AB loop exhibited systematic contacts with the surface. This AB-loop is shown to be only loosely bound during the simulation see

Fig. 3(c) and it can detach itself from the surface. On the contrary, in Fig. 3(a) the contact patch through N-TER and DE loop is well conserved during the entire 20 ns length of T-REMD and the protein remains anchored through the N-terminal residues (ILE1, GLN2, ARG3) and DE-loop residues (LYS58, TRP60).

T-REMD results revealed that D76N mutant, if compared to the wild-type β2m, is characterised by a greater flexibility along the AB loop (res 12–20) and EF loop (res 71–77), the latter containing the mutated residue 76. Due to the substitution of ASP76 with ASN76, the donor/acceptor atoms belonging to the neighbouring AB and EF loops are disrupting salt bridges allowing the formation of essential hydrogen bonds. As a result, a large degree of detaching of AB loop from EF loop is observed during the dynamics.

Loop AB, however, showed poor or no involvement at all in the NMR monitored samples of D76N, except for the attenuation of GLU16 or HIS13, as will be discussed in the Experimental part. Thus, the association of D76N into dimers and their direct interaction with cit-AuNP is investigated in the following sections (see section "Protein–protein docking: prediction of dimeric interface"), looking for a better comparison with available experimental data.

For ΔN6 variant, in the lower panel of Fig. 3, the contact patch identified by docking is confirmed to be well conserved, since the protein remains anchored through the DE-loop residues (LYS58, TRP60) and BC-loop residue (HIS31) during the entire 30 ns lenght of T-REMD. An additional contact is found through C-TER only in the last 20 ns of simulation. The capability of the ΔN6 protein to remain anchored to the citrate surface during T-REMD and the partial involvement of C-TER region is in line with the behavior of native protein, and in good agreement with the experimental data, as it will be discussed in the "Experimental part".

## Protein–protein docking: the dimeric interface

In order to provide a complementary approach to the interpretation of available experimental data, a preliminary docking to

**Table 3** Resultant cit-AuNP-dimer encounter complexes from rigid-body BD docking of D76N$_{chg}^{net}$ = +6.00e and ΔN6$_{chg}^{net}$ = −4.00e to a negative Au(111) surface. A hierarchical clustering algorithm (based on a minimum distance linkage function) was applied to the diffusional encounter complexes after docking to a bare negative gold (Au$_{chg}^{neg}$ = −0.05e) surface. The reported complexes represent for (D76N)$_2$-AuNP the 99.9% of the encounter complexes obtained by BD simulation, respectively and for (ΔN6)$_2$-AuNP the 99%

| Label | RelPop %[a] | $U_{Repr}$[b] | $E_{LJ}$[c] | $E_{LJ} + U_{ds}^p + U_{ds}^{m}$[d] | $U_{EP}$[e] | Contact residues[f] |
|---|---|---|---|---|---|---|
| **D76N** | | | | | | |
| 1-Ad1 | 99.9 | −44.22 | −4.56 | −2.153 | −42.07 | SER57 LYS58 MET99 |
| 1-Ad4 | 99.9 | −65.15 | −5.32 | −3.72 | −68.86 | N-TER LYS58 TRP60 |
| **ΔN6** | | | | | | |
| 1-Id1 | 99 | −5.5 | −11.02 | −6.50 | 1.0 | LYS58 TRP60 |
| 1-Id3 | 99 | −10.66 | −24.52 | −22.22 | 11.56 | LYS58 ASP59 TRP60 |

[a] Relative population of this cluster. [b] $U_{Repr}$: total interaction energy of the representative of the given cluster in kT with $T$ = 300 K. [c] $E_{LJ}$: Lennard-Jones energy term for the representative complex, $U_{ds}^p$: non-polar (hydrophobic) desolvation energy of the representative complex, in kT. [d] $U_{ds}^m$: surface desolvation energy of the representative complex, in kT. [e] $U_{EP}$: total electrostatic energy of the representative complex, in kT. [f] Residues with atoms contacting gold at distances ≤3 Å.

D76N dimer

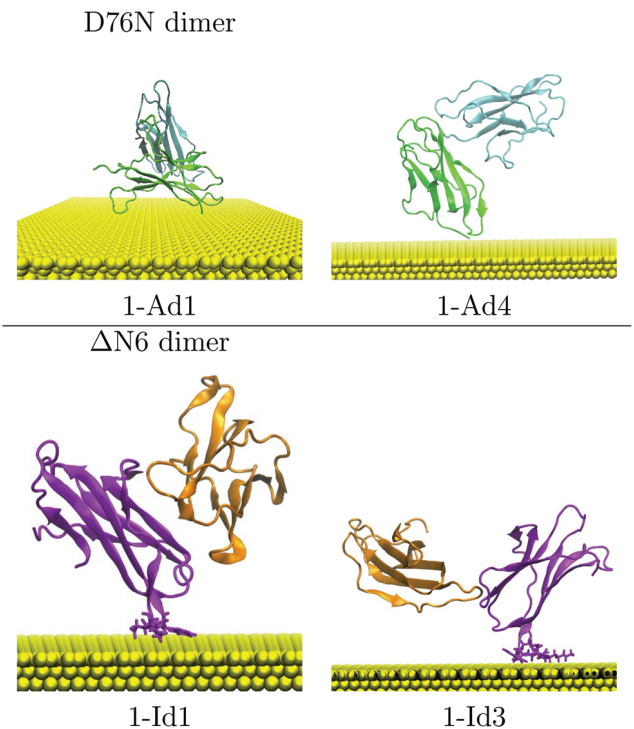

ΔN6 dimer

**Fig. 5** Most populated encounter complexes of dimeric $D76N_{chg}^{net}$ = +6.00e and $\Delta N6_{chg}^{net}$ = −4.00e on negatively charged gold nanocluster ($Au_{chg}^{neg}$ = −0.05e) obtained by BD simulation. The protein backbone is shown in cartoon representation. The residues contacting the gold surface are shown in stick representation.

build D76N dimers is performed. For the sake of completeness, ΔN6 dimer are also considered. The association of wild-type β₂m or variants into dimers and, to reduced extents, larger oligomers[28–31] in solution has been frequently observed. Here, rigid-body docking method implemented in SDA 7.2 are applied to predict the dimeric interfaces of the modified encounter complexes, which are supposed to exist pre-assembled in solution before the addition of cit-AuNP.

We wish to remark that the protein–protein Brownian Dynamics is performed as an initial sampling stage of protein–protein diffusional association in the presence of implicit solvent. The docked configurations obtained at this stage are then grouped with a hierarchical clustering algorithm into ensembles that represent potential protein–protein encounter complexes. Flexible refinement of selected representative structures is thus done by molecular dynamics (MD) simulations in explicit solvent. The advantage of using Brownian Dynamics is that it mimics efficiently the physical process of diffusional association of the unbound proteins whereas the atomistic refinement is accounting for the protein conformational flexibility upon association.

The protein–protein docking reported in this section and the MD refinements reported in the next section, represents a preliminary step towards the docking of D76N and ΔN6 dimeric adducts on cit-AuNP.

The adsorption free energies of the protein–protein encounter complexes of D76N–D76N and ΔN6–ΔN6 are reported in Table 2 along with the clustered trajectories.

**D76N dimers.** Docking to build proteins dimers starting from two identical D76N monomers was applied and it provided one main orientation accounting for more than 97% of all the protein–protein encounter complexes, as reported in Table 2. The representative structure of the most relevant complex is shown in the last column of the same Table 2. The contact residues are different for the two monomers (i.e. sub-units) forming the dimer.

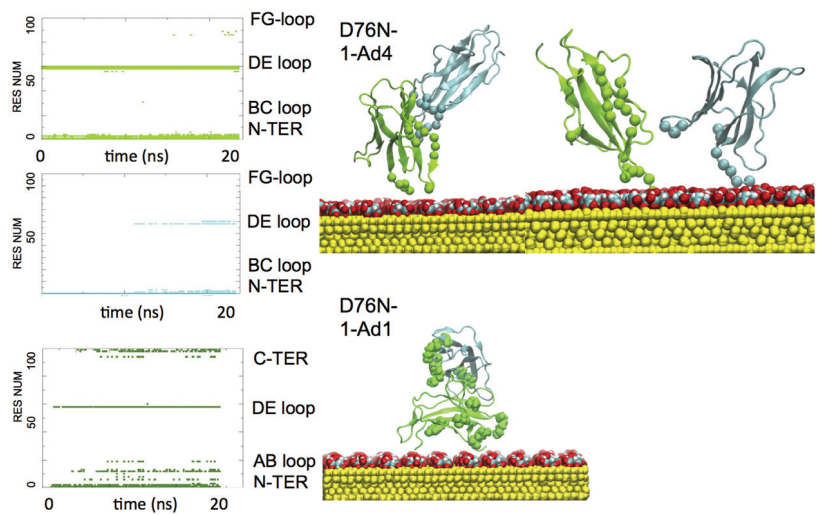

**Fig. 6** Top panel: (on the left) Time evolutions of D76N dimers contacting residues respect to the surface of the cit-AuNPs along the entire TREMD dynamics (i.e. protein residues within 3 Å from the surface). The binding patches established by each sub-unit with the cit-AuNP are differentiated by color (green for sub-unit 1 and cyan for sub-unit 2) (on the right). Most stable orientations of the 1-Ad4 dimer of d76N interacting with cit-AuNP. Direct contacts of the sub-unit 1 (green) and sub-unit 2 (cyan), are highlighted with balls on the α carbon atoms. Bottom panel: (left and right) The same representation is reported for second dimeric complex of D76N, namely 1-Ad1, interacting with cit-AuNP.

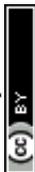

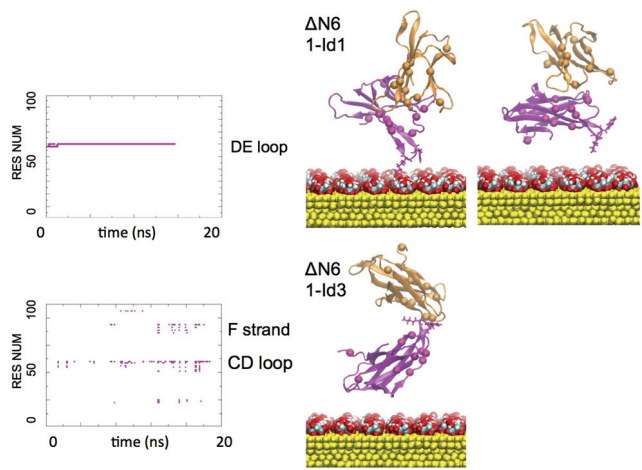

**Fig. 7** Top panel: (on the left) Time evolutions of ΔN6 dimers contacting residues respect to the surface of the cit-AuNPs along the entire TREMD dynamics (i.e. protein residues within 3 Å from the surface). The binding patches established by each sub-unit with the cit-AuNP are differentiated by color (magenta for sub-unit 1 and orange for sub-unit 2) (on the right). Most stable orientations of the 1-Id1 dimer of ΔN6 interacting with cit-AuNP. Direct contacts are occurring only through the subunit 1 (magenta) and they are highlighted with balls on the α carbon atoms. Bottom panel: (left and right) The same representation is reported for second dimeric complex 1-Id3 of ΔN6 interacting with cit-AuNP.

Sub-unit 1, depicted in cyan, involves in the dimeric interface the binding N-terminal (THR4 PRO5), BC loop (HIS31) and FG loop (THR86 SER88). Sub-unit 2, depicted in green, shows B strand (PHE22) CD loop, D strand (ILE46, GLU47, LYS48, GLU50, ASP53) and E strand (TYR67, GLU69) as interacting residues.

In the case of D76N–D76N dimers, the binding is driven both by Lennard-Jones and electrostatic interactions.

**ΔN6 dimers.** The docking provided two different orientations accounting for 70 and 30 per cent of the encounter complexes, respectively. The most stable and populated complex has a residue interface for sub-unit 1 (orange) involving SER33, TRP60, PHE62, LEU54 and sub-unit 2 (magenta) involving HIS31, PRO32, ASP34, THR86.

The most representative and most populated complexes for each systems, are shown in Table 2.

The results show that D76N variant has a more favourable attraction between monomers that facilitates aggregation with respect to ΔN6 (and native protein), at pH around neutrality. This can be interpreted as a consequence of the asparagine substitution for aspartate which has a substantial impact in the variant protein, despite the survived interaction between residues 42 and 76.[28,29]

For the sake of completeness, the stability of the protein–protein dimers has been examined using 400 ns of standard

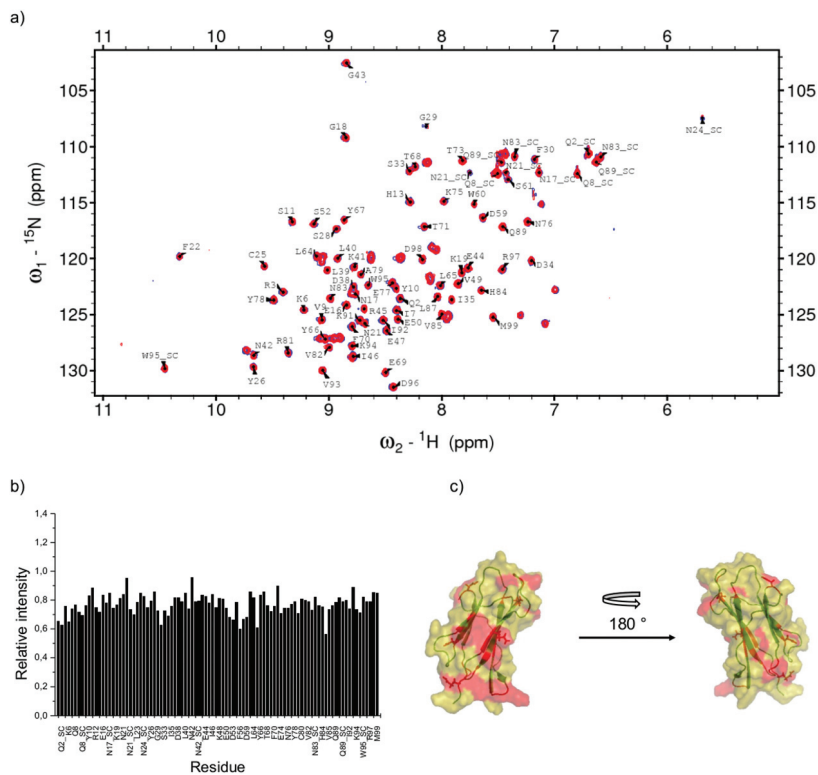

**Fig. 8** D76N β2m NMR results. (a) Overlay of 15N–1H HSQC spectra of 15N-labelled D76N β2m 18 microM in the free form in blue and in the presence of 90 nM cit-AuNP. (b) Bar plot of relative intensity calculated from the comparison between the spectra reported in (a). (c) D76N β2m cartoon highlighting the residue locations that proved most affected by cit-AuNPs. i.e. displaced one standard deviation at least with respect to the average relative intensity.

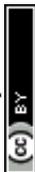

MD simulations in solution, see Fig. 4, starting with the most representative dimeric complexes obtained from rigid-body BD docking. Simulations were repeated four times using a different seed for the initial velocity distribution (d1, d2, d3, d4) for each system to improve the statistics of the search of the energy minima on the potential energy surface. Only the final most stable dimeric complexes are reported (more details are reported in the ESI†).

### Dimers on cit-AuNP

**Docking of dimers to negative gold.** The docking procedure is thus applied to the dimers on negative gold surface. From MD simulations in Fig. 4, two different dimers are obtained for each variant. More specifically, complexes A-d1, A-d4 pertain to D76N dimers and complexes I-d1 and I-d3 to ΔN6 dimers.

Docking results in Table 3 indicate that the surface charge has a crucial influence on the binding of the dimeric complexes on the negative AuNP. The electrostatic interactions play an important role in changing the relative stability of the most populated and stable complexes.

From Fig. 5, complex 1-Ad1 of D76N is stabilised *via* LYS58 residue and the interacting patch is characterised by the presence of both NTER and CTER (MET99) close to the Au surface. Complex 1-Ad4 of D76N is also stabilised *via* LYS58 residue and it is shifting the NTER towards the negative surface. Both D76N dimers ($\Delta N6^{net}_{chg}$ = +6.00$e$) benefits from a favourable electrostatic interactions with the negatively charged surface.

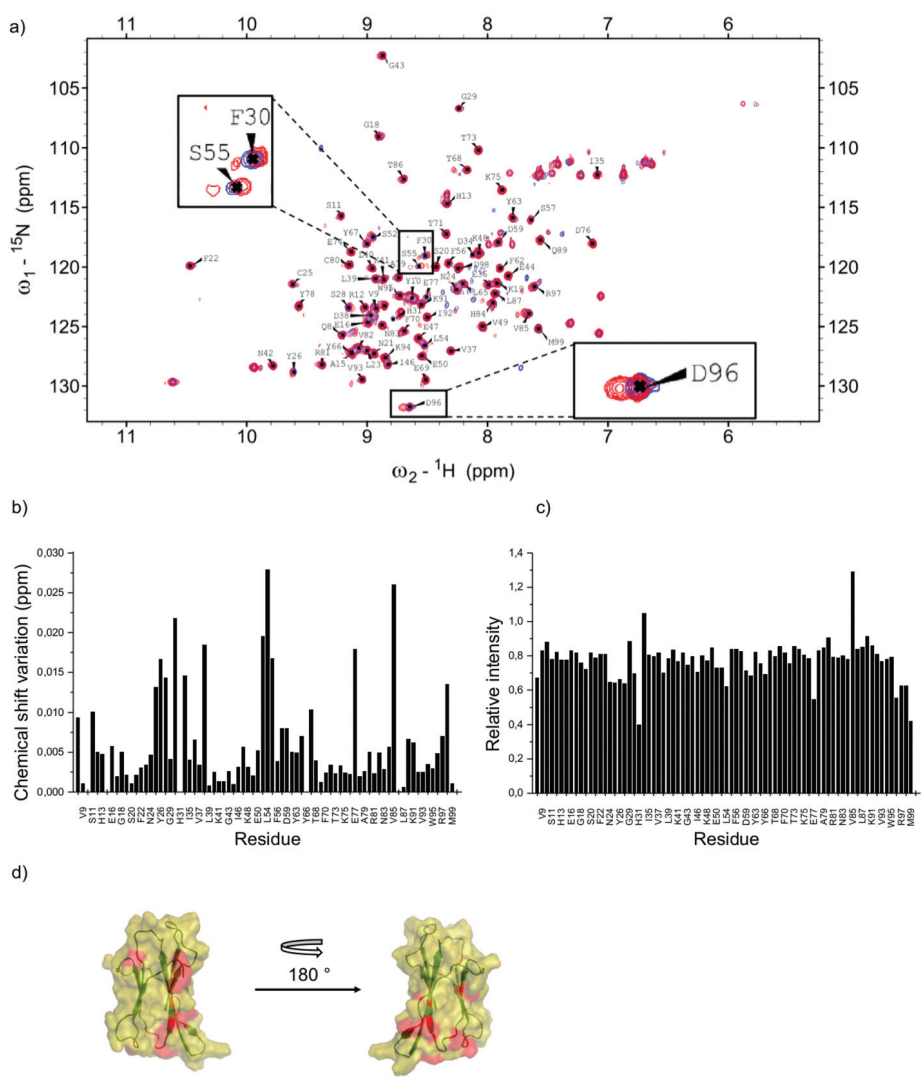

**Fig. 9** ΔN6 β₂m NMR results. (a) Overlay of $^{15}$N–$^{1}$H HSQC spectra of $^{15}$N-labelled ΔN6 β₂m 17 μM in the free form in blue and in the presence of 30 nM cit-AuNP. In the inserts an example of two chemical shift variations (F30 and S55) and the doubling of D96 signal are presented. (b) and (c) Bar plots of chemical shift variations and relative intensity, respectively, calculated from the comparison between the spectra reported in (a). (d) ΔN6 β₂m cartoon highlighting the residue locations that proved most affected by cit-AuNPs. *i.e.* displaced one standard deviation at least with respect to the average chemical shift variation.

On the contrary, ΔN6 dimers are both accompanied by an unfavourable electrostatic contribution due to the total negative charge of dimer ($\Delta N6_{chg}^{net} = -4.00e$). Resulting complexes 1-Id1 and 1-Id3 of ΔN6 interact textivia the LYS58 and TRP60 and LYS58, ASP59, TRP60, respectively. Complex 1-Id3 is accompanied by a more unfavourable electrostatic energy which is partially compensated by the LJ interaction.

**Enhanced sampling of dimers on cit-AuNP.** As a final step of the computational strategy, T-REMD simulations are applied to refine the interactions between the dimers of D76N (complexes 1-Ad4 and 1-Ad1) and of ΔN6 (complexes 1-Id1 and 1-Id3) on cit-AuNP.

Our simulations illustrate that the dimers of the two variants display a distinct behaviour towards the negatively charged surface of the cit-AuNP, due to their different total charges. The initial contact of both dimers onto the surface of the citrate layer is facilitated by Coulomb interactions between the positively charged residues at N-TER and/or loop DE (LYS58) and the oxygen anions of the citrate molecules. However, once protein flexibility is introduced, this molecular picture changes as the competition between protein–protein and protein–cit-AuNP interaction depends on electrostatics.

We found that the interaction with D76N dimers with cit-AuNP leads to complete dissociation of the dimeric adducts,[32] whereas for ΔN6 dimers the dissociation cannot be seen at the time length of the simulation. The gold–dimer interface of ΔN6 is found to be labile with respect to its gold–monomer interface and also to the gold–dimer interface of D76N. The electric field created by the cit-AuNP, in fact, is not strong enough to prevent the formation of stable complexes with ΔN6 monomers ($\Delta N6_{chg}^{net} = -1.00e$) but it weakens the interaction with dimers carrying a larger negative charge ($\Delta N6_{chg}^{net} = -4.00e$) due to an enhanced protonation state after dimerisation as explained in Methodology, see Fig. 7.

**D76N 1-Ad4.** Results reported in Fig. 6, account for the protein approaching the cit-AuNPs at the N-terminal tail and at the DE loop of the single sub-unit 2 (green) within D76N dimer. The interaction between the protein and the AuNP surface exhibits an initial state where just the sub-unit 2 (green) is involved in the vicinity of the surface. However, after running T-REMD, the final state displays a configuration where the dimer is essentially disassembled[32] (see Fig. 6, with sub-unit-1 and sub-unit 2 interacting with the citrate layer through the N-terminal fragment or the DE loop).

The crucial points of this result is the significance of the interaction with cit-AuNPs that essentially leads to the complete dissociation of the dimer, *i.e.* disruption of the very first step of aggregation.

**D76N 1-Ad1.** For the sake of completeness, docking with the "zipped" dimer is reported, showing a direct contact with cit-AuNP involving the unique sub-unit 2, see Fig. 6. Sub-unit 2 touches the cit-AuNP surface through N-TER, AB loop (res 1, 3, 12, 13, 19) while DE loop (LYS58) and CTER (93, 97, 99) residues of the same sub-unit contact the surface only upon structural relaxation at the interface. Results are reported in Fig. 6.

Interestingly, the binding patch of sub-unit 2 with cit-AuNP is identical to the binding patch of sub-unit 2 with sub-unit 1 (see Fig. 4), suggesting that the interaction of this dimer with the cit-AuNPs is potentially able to block active sites of one monomer for the binding to another protein, inhibiting the growth of further protein–protein interactions. Given the more extended protein/protein dimeric interface at the zipped dimer, the detachment of sub-unit 1 from sub-unit 2 for this complex, is not seen at the time length of the simulation but an overall weakening of the protein–protein interface is observed as a consequence of salt-bridges breaking (see ESI†).

**Table 4** Direct comparison between experimental chemical shift deviations and the computed contacting residues of D76N (top) and ΔN6 (bottom) monomers and dimers at the protein–NP interface from T-REMD refinement

| Structure region | NMR attenuations | Comp. monomer | Comp. dimer | Comp. dimer |
|---|---|---|---|---|
| D76N | | D76N-A | 1-Ad1 (zipped) | 1-Ad4 (unzipped) |
| N-ter, A strand | 2sc, 6, 7, 8sc | 1, 3, 6 | 1, 3 | 4, 5 |
| AB loop | 13, 16, 17 | 11,12,19 | 12, 13, 19 | |
| B strand | 21sc, 24sc, 28 | | 26 | 22 |
| BC loop | 30, 33, 34 | | | 31 |
| CC′, C′D loops | 42sc, 43 | | 40 | 46, 47, 48 |
| D strand | | | | 53 |
| DE loop | 58, 59 | 58, 60 | 58 | |
| E strand | 64, 65, 66, 70 | | 75 | 67, 69 |
| F strand | 83sc | | | |
| FG loop | | | | 86, 88 |
| G strand, C-ter | 91, 93, 95sc, 97 | | 93, 97, 99 | |

| Structure region | NMR chemical | Comp. monomer | Comp. dimer | Comp. dimer |
|---|---|---|---|---|
| ΔN6 | shifts | ΔN6-I | 1-Id1 (zipped) | 1-Id3 (unzipped) |
| N-ter, A strand | | | | |
| AB loop | | 8, 10, 11,12,19 | | 19, 20 |
| B strand | 25, 26, 28 | | | 41 |
| BC loop | 30, 31, 34, 38 | 30, 31 | | |
| CC′, C′D loops | | | | 47, 48 |
| D strand | 52, 54, 55 | | | |
| DE loop | | 58, 60 | 58, 59, 60 | |
| E strand | | | | |
| F strand | 77 | | | 75 |
| FG loop | 85 | 88 | | |
| G strand, C-ter | 98, 99 | 97, 98, 99 | | |

**ΔN6 1-Id1, 1-Id3.** Results illustrate that in the presence of ΔN6 dimers, the protein–protein AuNPs interface is more labile at a physiologically relevant pH, as a consequence, 1-Id1 (zipped) dimer can attach and detach from the surface during TREMD but several contacts are observed. We wish to remark that the dimeric interface of (zipped) dimer with cit-AuNP is indistinguishable from the monomeric interface (res 58, 59, 60). In fact, the direct interaction with the surface is occurring *via* a single sub-unit, as shown in Fig. 7. This suggests that for this species the dominant interaction may occur both with the monomers and with dimer (*i.e.* the monomer within the dimer), providing the same binding interface. Also in this case, the complete detachment of sub-unit1 from sub-unit 2, is not seen at the time length of the simulation but a change in the protein–protein interface leading to hydrogen bond breaking is seen during the simulations.

In the case of 1-Id3 (unzipped) it is not possible to identify a really stable binding patch for ΔN6 dimers at the time length of our simulations but only some unstable contact with positively charged residues (*e.g.* LYS48 and LYS75).

## NMR experimental evidence

To map experimentally the interaction of the two β₂m variants tested in the simulations, $^{15}$N–$^{1}$H HSQC spectra of the two proteins without and with cit-AuNPs were collected using protein/NP ratios of 213 and 567 for D76N and ΔN6, respectively. In spite of the rather conspicuous ratio difference, the examined solutions had approximately equal protein concentrations (the NP concentration was around 90 or 30 nM), which rules out artifacts due to the actual behaviour of the proteins reflecting essentially the amount of free species. The effect of the nanoparticle presence on the intensity and position of nitrogen–hydrogen correlation peaks was assessed. As we reported before,[32] for D76N variant we observed no chemical shift variation but an intensity decrease with an average intensity ratio between the signals in the two spectra of 0.78 ± 0.04. The relative intensity profile shows a differential pattern indicating that there are specific residues preferentially affected by the presence of cit-AuNP. Concerning ΔN6 variant, in addition to intensity decrease, resulting in an average value of 0.78 ± 0.12, some peaks undergo also slight change in their chemical shift (see Fig. 8 and 9). Other two features can be noticed in ΔN6 HSQC spectrum recorded in presence of cit-AuNP: the doubling of D96 peak and the intensity increase of two peaks, namely V85 and D34. Similar effects were observed also when ΔN6 was monitored at lower concentration (4 μM, not shown), with the same cit-AuNP preparation. In general, intensity and chemical shift changes may report either the protein inter-action surface with cit-AuNP and/or the protein–protein interaction evolution in the presence of cit-AuNP, as observed with D76N.[32] Besides the direct contact effects, any such interactions may prove capable of altering local conformations, which also may lead to intensity and/or chemical shift deviations. This must then be the case also with the changes in ΔN6 spectra. The clustering pattern of the chemical shift and intensity deviations depicted in Fig. 8 and 9, along with the previously reported results for wild-type[9] and D76N,[32] suggest

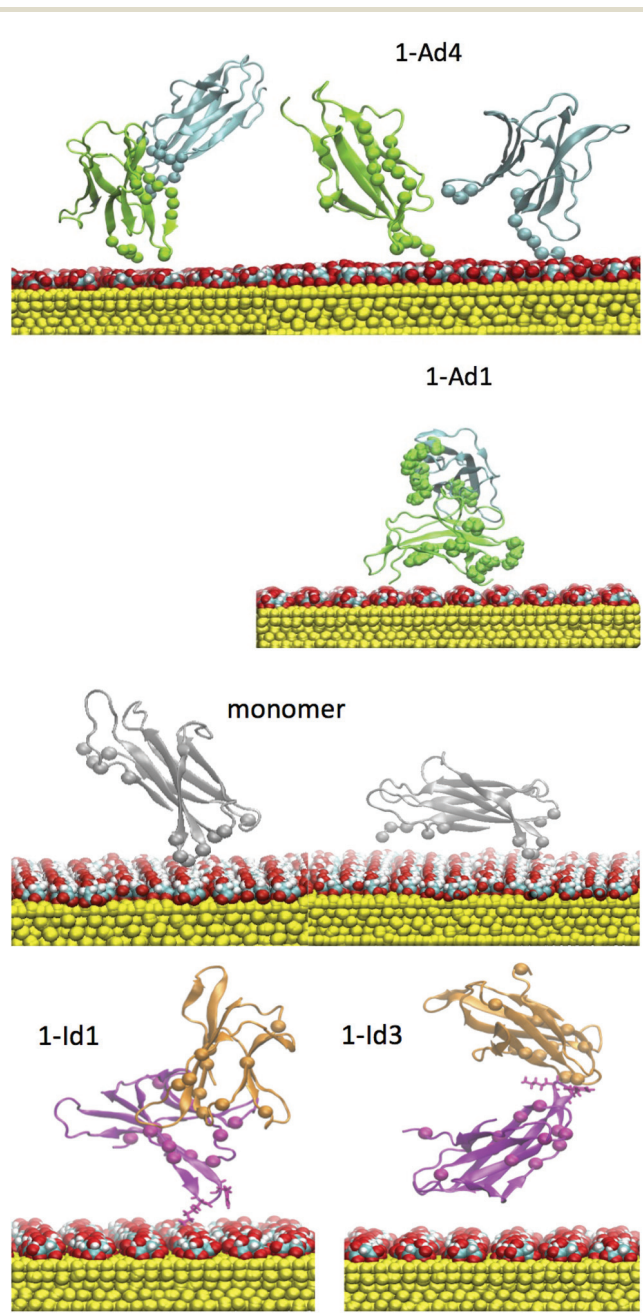

**Fig. 10** Final orientations resulted from T-REMD refinement of D76N and ΔN6 on cit-AuNP. Dimeric complexes 1-Ad4 and 1-Ad1 are obtained for D76N. Monomeric complex and 1-Id1, 1-Id3 dimeric complexes are obtained for ΔN6.

that a typical protein/cit-AuNP contact interface encompasses the N-terminal surrounding apical region (BC, DE and FG loops along with the N-terminal stretch where present). The additional involvements should reflect changes due to the shift of the protein/protein association equilibria that are elicited by competing protein/NP interactions and may concern either the association interface or allosteric structural effects that propagate to buried regions. All of these effects were already recognised in D76N/cit-AuNP systems[32] and seem to occur also with ΔN6, with appreciable effects also on chemical shifts. In particular, in Table 4 and Fig. 10 we report a direct comparison between experimental chemical shift deviations and the contacting residues of D76N and ΔN6 variants at the protein–NP interface. Our simulations both reproduce and explain the experimentally observed data. The deviations observed at B-strand, F-strand and C-terminal fragment of ΔN6 may feature a decrease of association, with consequent local rearrangements at strand B, in the presence of cit-AuNP. The cross-peak splitting distinctly observed for D96 amide signal is likely to arise from a decreased inter-conversion rate of two limiting local conformers following the association pattern change. The simulation results support the establishment of different conformation of the C-terminal region upon interaction with the cit-AuNP, as shown in Fig. 11. The plot of the distance between D96 residue and the neighbouring V9 show the presence of three distinct peaks associated to different conformers, namely conformer α in solvent and conformers β and γ with cit-AuNP, in which the C-terminal tail is close/distant to strand A (V9). The cross-peak splitting observed experimentally could then reflect either the inter-conversion between the most populated peak α in solvent and peak γ on cit-AuNP or between conformers β and γ. Since those conformers β and γ are not observed in solvent, their onset can be ascribed to the interaction with the cit-AuNP surface. On the other hand, the intensity increase coupled to resonance shift of V85 and D34 amide cross-peaks could reflect local decrease of dipolar or/and exchange broadening arising from cit-AuNP contact, associated to a conformational change at BC and DE loop as resulted from simulations (Fig. 12). The conformation-

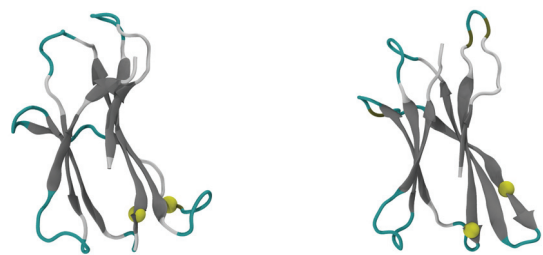

V85 and D34 in solvent    V85 and D34 on cit-AuNP

**Fig. 12** ΔN6 variant: residues V85 and D34 (yellow dots) undergo a different conformation of BC loop and DE loop going from solvent to cit-AuNP.

al changes of those loops are known to play an important role in fibrillation process.[33]

### Diffusion coefficient determinations

NMR 2D DOSY spectra[34] were collected to measure the translational diffusion coefficients of the β₂m variants in the absence and presence of cit AuNPs. The results clearly show that the diffusion coefficients of the proteins increase when cit AuNPs are present in solution Fig. 13. This is consistent with an effect of cit AuNP on the protein association equilibria that prove all shifted towards the monomeric species.

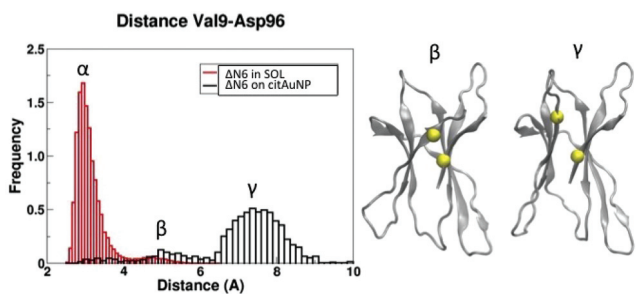

**Fig. 11** ΔN6 variant: the plot report the distance between D96 residue and the neighbouring V9, in solvent (red plot) and for the protein interacting with cit-AuNP (black plot). Results are obtained analysing the most populated structures after clustering. The figure show the presence of two distinct peaks associated to different conformers in which the C-terminal tail is bound/unbound to strand A.

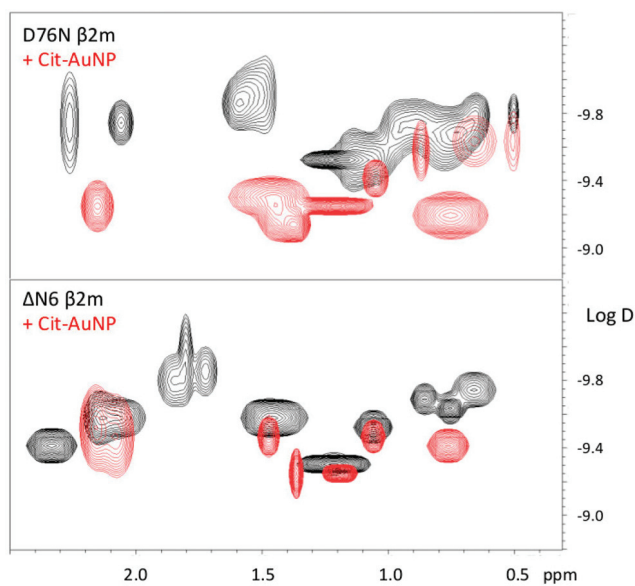

**Fig. 13** DOSY map overlays of D76N and ΔN6 β₂m without (black) and with (red) cit-AuNP. Proteins were always 4 μM and cit-AuNP was 20–30 nM and measurements were run at 25 °C. An increased diffusion coefficient (D in m² s⁻¹) is observed in the presence of cit-AuNP. The overwhelming signals from Hepes buffer and citrate, in addition to the residual solvent peak, restrict the region where isolated protein peaks can be observed.

# Conclusions

In conclusion, we have proposed a comprehensive and consistent mechanism for how citrate covered gold NP influence protein aggregation (dimerization) and thus fibril formation for the highly amyloidogenic variants D76N and ΔN6. The characterisation of the interaction between cit-AuNPs and D76N, ΔN6 β2m naturally occurring variants, by atomistic simulations has shed light on the microscopic mechanism of the process. By NMR we have mapped preferential interaction sites on the protein surface that have been properly reproduced by modelling calculations. We propose that the prominent dimer population of D76N at neutral pH undergoes efficient reshaping and eventually splitting in the presence of a cit-AuNP surface. Conversely, the similar distribution of association adducts of ΔN6 protein interacts less efficiently, *i.e.* with reduced turnover frequency, with cit-AuNPs. The reduced efficiency accompanies the electrostatic repulsion contributions that are estimated more unfavourable for the dimer with respect to monomers. While in simulations this effect determines indistinct extent of interaction of monomers and dimers with cit-AuNPs, the experimental NMR pattern of ΔN6 shows more pronounced consequences on the chemical shifts, with respect to the corresponding pattern observed with D76N. For both proteins, however, the NMR evidence also demonstrates consistently a reduction of the association extent in presence of nanoparticles as inferred from the increase of the translational diffusion coefficient. In some of the simulated systems, we found that the interaction of D76N dimers with cit-AuNP leads to complete dissociation of the dimeric adducts, whereas for ΔN6 dimers the dissociation could not be seen at the time length of the simulation. However, the interaction with cit-AuNPs is always seen to interfere at the sites of protein–protein interaction and to lead, conceivably, to an inhibition of the fibrillation events.

## Methodology

**Brownian dynamics simulations.** Rigid-body docking simulations were carried out using Brownian dynamics (BD) techniques with the ProMetCS continuum solvent model for protein–gold surface interactions.[16,20] The calculations were performed using the SDA version 7 software.[17] The β2m structure was taken from the NMR solution structure (PDB id: 1JNJ) and the mutation at residue 76 was introduced manually, as the truncation of the first six residues of ΔN6.

Titratable protein side chains, were assigned at pH 7.2 with H++.[18] As in ref. 9, in addition to HIS51 and HIS84 even HIS31 is protonated, given the presence of the negative citrate adlayer which may stabilise the protonated regime. For dimers the protonation state was assigned after dimerisation in explicit solvent. We wish to remark that the ΔN6 monomers in solvent has an initial $\Delta N6_{chg}^{net} = -1.00e$ but after dimerisation the protonation state of each monomeric sub-unit becomes $\Delta N6_{chg}^{net} = -2.00e$.

5000 BD trajectories were computed starting with the proteins positioned randomly with its center at a distance of 70 Å from the surface where the protein–surface interaction energy is negligible. The specified number of docked complexes was extracted directly from the runs and clustered with a clustering algorithm. The relative translational diffusion coefficient was 0.0123 $Å^2$ $ps^{-1}$ and the rotational diffusion coefficient for the protein was $1.36 \times 10^{-4}$ in $radian^2$ $ps^{-1}$. The simulation time step was set to 0.50 ps. Parameters for the calculation of hydrophobic desolvation energy/forces was set to $-0.019$ kcal $mol^{-1}$ $Å^{-2}$ and for the electrostatic desolvation energy/forces to 1.67 according to ref. 19. BD trajectories were generated in a rectangular box (ibox = 1); the dimensions of the $(x, y)$ plane, describing the symmetry of the simulation volume as well as the surface size, were given as input parameters. At each BD step, the protein–protein interaction energy and forces acting on the protein were computed using the implicit-solvent ProMetCS forcefield,[20] developed and parametrised for protein–gold surface interactions. The energy terms included in ProMetCS have been described in the main text.

We applied a single-linkage clustering method (based on CA atoms, with RMSD = 3.0 Å) algorithm and parameters providing the smallest number of physically distinct orientations of β2m on cit-AuNP, for all the results given in the manuscript.

**Molecular dynamics simulations.** We used our own force field parameters for the citrate anions based on *ab initio* calculations. The same protein and gold structures as for the BD simulations were used for the initial coordinates for the MD simulations. A rectangular simulation box of dimensions (101.5 Å × 99.6 Å × 101.5 Å) including SPC/E water molecules, the protein monomers and dimers and the gold surface was built. The protein was placed at the positions of the representatives of the docked clusters obtained from the BD docking simulations. Before the addition of the water molecules, the center of mass of the protein was placed at 47 Å from the surface, retaining the original docked orientation with respect to the surface. The choice of this distance was motivated by various tests that we performed showing that if the simulations were started with the protein in direct contact to the surface (or at smaller distances), it was in a kinetically trapped state where only minor relaxation could take place on the timescale of tens of ns. During equilibration dynamics, all systems contacted the surface within the first 1 ns of MD without reorienting respect to the surface.

All simulations were performed with the Gromacs 5.2.1 package.[25] GolP[16] and OPLS/AA parameters[26] were used for the surface and the protein and the SPC/E water model[27] was applied. The lengths of bonds were constrained with the LINCS algorithm. Surface gold atoms and bulk gold atoms were frozen during all simulations but gold dipole charges were left free. Classical MD simulations were performed at constant volume and temperature ($T = 300$ K). Periodic boundary conditions and the Particle-Mesh-Ewald algorithm were used. A 2 fs integration time step was used. For the citrate anions we have implemented new force field parameters based on *ab initio* calculations (that take into account the quantum nature of such small chemical species) in a consistent and compatible way with the existing GolP force field for the protein–AuNP surface interactions.

We worked out an enhanced MD sampling performed with T-REMD (Temperature Replica Exchange) simulations in explicit water on the most relevant encounter complex found along the docking. The sampling was enhanced by introducing temperature swapping moves between states with similar density at different temperatures. We employ a total of 32 replicas, covering the temperature range between 290 and 320 K.

Principal component analysis, clustering analysis, hydrogen bond and salt bridges analysis were also performed using GROMACS.

**NMR.** $^{15}$N-labelled D76N and $\Delta$N6 $\beta_2$m solutions in 25 mM phosphate buffer and 50 mM HEPES, respectively, at pH 7 were analysed with and without cit-AuNPs by recording 2D $^{15}$N–$^1$H HSQC.[35] Spectra were collected at 14.0 T, on the Bruker Avance III NMR facility of the Core Technology Platform at New York University Abu Dhabi. The spectrometer, equipped with cryoprobe and z-axis gradient unit, operated at 600.13 and 60.85 MHz to observe $^1$H and $^{15}$N, respectively. Spectral widths of 40 ppm ($^{15}$N, t1) and 15 ppm ($^1$H, t2) were used. For each t1 dimension point, 128 or 64 scans were accumulated and quadrature in the same dimension was accomplished by gradient-assisted coherence selection (echo-antiecho).[36] Processing with t1 linear prediction, apodization and zero-filling prior to Fourier transformation led to 2K1K real spectra. Water suppression was achieved by using a flip-back pulse in the HSQC experiments.[37] All measurements were performed at 25 C. Spectra were processed with Topspin 2.1 and analysed with Sparky.[38] Chemical shift deviations were calculated as $\Delta\delta$ (ppm) = $[(\Delta\delta H)2 + (\Delta\delta N/6.5)2]1/2$ where $\Delta\delta$ H and $\Delta\delta$ N are the chemical shift variations for $^1$H and $^{15}$N, respectively,[39] whereas the relative intensity is the ratio of the peak intensity in the presence of cit-AuNP and in the absence.

Diffusion coefficients were determined by means of 2D $^1$H DSTEBPP (Double STimulated Echo BiPolar Pulse) experiments.[40] Protein concentration was 4 μM in 50 mM Hepes, pH = 7 in 95/5 $H_2O/D_2O$, either in absence and in presence of cit-AuNP. Sodium citrate (1.5 mM) was present in the absence of NP. The z-axis gradient strength was varied linearly from 10 to 90% of its maximum value (~60 G cm$^{-1}$) and matrices of 2048 by 40 points were collected by accumulating 512 scans per gradient increment. Water suppression was carefully adjusted by appending to the DSTEBPP sequence a pair of WATERGATE[41] elements in the excitation-sculpting mode.[42] Careful setting was the acquired data were processed using the Bruker software Dynamics Center to extract the diffusion coefficients.

## Conflicts of interest

There are no conflicts to declare.

## Acknowledgements

Funding from MIUR through PRIN 2012A7LMS3_003 is gratefully acknowledged. S. C. acknowledges funding from ERC under the grant ERC-CoG-681285 TAME-Plasmons. The ISCRA staff at CINECA (Bologna, Italy) is acknowledged for computational facilities and technical support. Oak Ridge National Laboratory by the Scientific User Facilities Division, Office of Basic Energy Sciences, U.S. Department of Energy is acknowledged for the supercomputing project CNMS2013-064. Facilities of the National Energy Research Scientific Computing Center (NERSC), which is supported by the Office of Science of the U.S. Department of Energy under Contract No. DE-AC02-05CH11231, are also acknowledged.

## References

1 X. R. Xia, N. A. Monteiro-Riviere and J. E. Riviere, *Nat. Nanotechnol.*, 2010, **5**, 671.

2 A. E. Nel, L. Mädler, D. Velegol, T. Xia, E. M. V. Hoek, P. Somasundaran, F. Klaessig, V. Castranova and M. Thompson, *Nat. Mater.*, 2009, **8**, 543.

3 D. F. Moyano and V. M. Rotello, *Langmuir*, 2011, **27**, 10376.

4 A. M. Gobin, E. M. Watkins, E. Quevedo, V. L. Colvin and J. L. West, *Small*, 2010, **6**, 745.

5 P. Wang, X. Wang, L. Wang, X. Hou, W. Liu and C. Chen, *Sci. Technol. Adv. Mater.*, 2015, **16**, 1.

6 N. Gilbert, *Nature*, 2009, **460**, 937.

7 J. A. Kim, A. Salvati, C. Aberg and K. A. Dawson, *Nanoscale*, 2014, **6**, 14180.

8 I. Lynch and K. A. Dawson, *Nano Today*, 2008, **3**, 40.

9 G. Brancolini, A. Corazza, M. Vuano, F. Fogolari, M. C. Mimmi, V. Bellotti, M. Stoppini, S. Corni and G. Esposito, *ACS Nano*, 2015, **9**, 2600.

10 F. Gejyo, T. Yamada, S. Odani, Y. Nakagawa, M. Arakawa, T. Kunitomo, H. Kataoka, M. Suzuki, Y. Hirasawa, T. Shirahama, *et al.*, *Biochem. Biophys. Res. Commun.*, 1985, **129**, 701.

11 S. Linse, C. Cabaleiro-Lago, W.-F. Xue, I. Lynch, S. Lindman, E. Thulin, S. E. Radford and K. A. Dawson, *Proc. Natl. Acad. Sci. U. S. A.*, 2007, **104**, 8691.

12 S. Valleix, J. D. Gillmore, F. Bridoux, P. P. Mangione, A. Dogan, B. Nedelec, M. Boimard, G. Touchard, J.-M. Goujon, C. Lacombe, P. Lozeron, D. Adams, C. Lacroix, T. Maisonobe, V. Planté-Bordeneuve, J. A. Vrana, J. D. Theis, S. Giorgetti, R. Porcari, S. Ricagno, *et al.*, *Engl. J. Med.*, 2012, **366**, 2276.

13 V. Bellotti, M. Gallieni, S. Giorgetti and D. Brancaccio, *Semin. Dial.*, 2001, **14**, 117.

14 T. Eichner, A. P. Kalverda, G. S. Thompson, S. W. Homans and S. E. Radford, *Mol. Cell*, 2011, **41**, 161.

15 P. Mangione, G. Esposito, A. Relini, S. Raimondi, R. Porcari, S. Giorgetti, A. Corazza, F. Fogolari, A. Penco, Y. Goto, Y.-H. Lee, H. Yagi, C. Cecconi, M. M. Naqvi, J. D. Gillmore, P. N. Hawkins, F. Chiti, R. Rolandi,

G. W. Taylor, M. B. Pepys, M. Stoppini and V. Bellotti, *J. Biol. Chem.*, 2013, **288**, 30917.

16 F. Iori, R. Di Felice, E. Molinari and S. Corni, *J. Comput. Chem.*, 2009, **30**, 1465.

17 M. Martinez, N. J. Bruce, J. Romanowska, D. B. Kokh, M. Ozboyaci, X. Yu, M. A. Äztürk, S. Richter and R. C. Wade, *J. Comput. Chem.*, 2015, **36**, 1631.

18 J. C. Gordon, J. B. Myers, T. Folta, V. Shoja, L. S. Heath and A. Onufriev, *Nucleic Acids Res.*, 2005, **33**, W368.

19 A. H. Elcock, R. R. Gabdoulline, R. C. Wade and J. A. McCammon, *J. Mol. Biol.*, 1999, **291**, 149.

20 D. B. Kokh, S. Corni, P. J. Winn, M. Hoefling, K. E. Gottschalk and R. C. Wade, *J. Chem. Theory Comput.*, 2010, **6**, 1753.

21 R. R. Gabdoulline and R. C. Wade, *J. Phys. Chem.*, 1996, **100**, 3868.

22 Y. Lin, G. Pan, G. J. Su, X. H. Fang, L. J. Wan and C. L. Bai, *Langmuir*, 2003, **19**, 10000.

23 J. Kunze, I. Burgess, R. Nichols, I. Buess-Herman and J. Lipkowski, *J. Electroanal. Chem.*, 2007, **599**, 147.

24 M. Hoefling, F. Iori, S. Corni and K. E. Gottschalk, *Langmuir*, 2010, **26**, 8347.

25 D. van der Spoel, E. Lindahl, B. Hess, G. Groenhof, A. E. Mark and H. J. C. Berendsen, *J. Comput. Chem.*, 2005, **26**, 1701.

26 W. L. Jorgensen, D. S. Maxwell and J. TiradoRives, *J. Am. Chem. Soc.*, 1996, **118**, 11225.

27 B. Hess and N. F. van der Vegt, *J. Phys. Chem. B*, 2006, **110**, 17616.

28 S. Giorgetti, S. Raimondi, K. Pagano, A. Relini, M. Bucciantini, A. Corazza, F. Fogolari, L. Codutti, M. Salmona, P. Mangione, L. Colombo, A. De Luigi, R. Porcari, A. Gliozzi, M. Stefani, G. Esposito, V. Bellotti and M. Stoppini, *J. Biol. Chem.*, 2011, **286**, 2121.

29 G. Esposito, R. Michelutti, G. Verdone, P. Viglino, H. Hernández, C. V. Robinson, A. Amoresano, F. Dal Piaz, M. Monti, P. Pucci, P. Mangione, M. Stoppini, G. Merlini, G. Ferri and V. Bellotti, *Protein Sci.*, 2000, **9**, 831.

30 D. Gumral, F. Fogolari, A. Corazza, P. Viglino, S. Giorgetti, M. Stoppini, V. Bellotti and G. Esposito, *Magn. Reson. Chem.*, 2013, **51**, 795.

31 G. Esposito, M. Garvey, V. Alverdi, F. Pettirossi, A. Corazza, F. Fogolari, M. Polano, P. P. Mangione, S. Giorgetti, M. Stoppini, A. Rekas, V. Bellotti, A. J. Heck and J. A. Carver, *J. Biol. Chem.*, 2013, **288**, 17844.

32 C. Cantarutti, S. Raimondi, G. Brancolini, A. Corazza, S. Giorgetti, M. Ballico, S. Zanini, G. Palmisano, P. Bertoncin, L. Marchese, P. Mangione, V. Bellotti, S. Corni, F. Fogolari and G. Esposito, *Nanoscale*, 2017, **9**, 3941.

33 G. Esposito, A. Corazza and V. Bellotti, *Subcell. Biochem.*, 2012, **65**, 1917.

34 K. F. Morris and C. S. Johnson Jr., *J. Am. Chem. Soc.*, 1992, **114**, 3130.

35 G. Bodenhausen and D. J. Ruben, *Chem. Phys. Lett.*, 1980, **69**, 185.

36 J. Keeler, R. T. Clowes, A. L. Davis and E. D. Laue, *Methods Enzymol.*, 1994, **239**, 145.

37 S. Grzesiek and A. Bax, *J. Am. Chem. Soc.*, 1993, **115**, 12593.

38 T. D. Goddard and D. G. Kneller, *SPARKY 3*, University of California, San Francisco.

39 F. A. A. Mulder, D. Schipper, R. Bott and R. Boelens, *J. Mol. Biol.*, 1999, **292**, 111.

40 A. Jerschow and N. Müller, *J. Magn. Reson.*, 1998, **132**, 13.

41 M. Piotto, V. Saudek and V. Sklenar, *J. Biomol. NMR*, 1992, **2**, 661.

42 T.-L. Hwang and A. J. Shaka, *J. Magn. Reson., Ser. A*, 1995, **112**, 275.